# Programming Soft Robots with Flexible Mechanical Metamaterials

Ahmad Rafsanjani[1], Katia Bertoldi[2,3,4,*], André R. Studart[1,*]

[1] Complex Materials, Department of Materials, ETH Zürich, 8093 Zürich, Switzerland,
[2] John A. Paulson School of Engineering and Applied Sciences, Harvard University, Cambridge, MA 02138, USA,
[3] Kavli Institute, Harvard University, Cambridge, MA 02138, USA
[4] Wyss Institute for Biologically Inspired Engineering, Cambridge, MA 02138, USA

*To whom correspondence should be addressed:
 bertoldi@seas.harvard.edu & andre.studart@mat.ethz.ch

**The complex behavior of highly deformable mechanical metamaterials can substantially enhance the performance of soft robots.**

Metamaterials are rapidly emerging from electromagnetic, acoustic, or mechanical properties governed by structure rather than composition. Mechanical metamaterials, in particular, have been designed to show superior mechanical properties, such as ultrahigh stiffness and strength-to-weight ratio, or unusual properties, such as a negative Poisson's ratio and a negative coefficient of thermal expansion. Whereas earlier research focused on designing mechanical metamaterials with linear elastic responses, more recently, nonlinear large deformations and mechanical instabilities - typically associated with failure - have emerged as promising tools for new functionalities, including programmable shape transformations, tunable mechanical properties, and energy absorption (1). Ongoing advances in additive manufacturing technologies facilitate the fabrication of functional mechanical metamaterials with unprecedented complexity.

Exploiting such complex metamaterials in advanced soft robots may lead to paradigm shifts in design, manufacture, and perception of future intelligent machines. Instead of assembling individual actuators (2), soft robot designers will be able to use conformable monolithic systems that can undergo complex motion directly programmed within the architecture of the mechanical metamaterial. Use of such programmable metamaterial architectures makes possible, for example, soft robots that can transform a simple input, such as a pressure impulse, into a complex sequence of flexion, tension, and torsion outputs. The ability to code morphing information in the metamaterial structure also allows such architected actuators to perform complex tasks using less input energy compared with their conventional counterparts.

## A Metamaterial Approach to Soft Robotics

Taking a metamaterial approach toward the design of soft machines substantially increases the number of degrees of freedom in deformation and the available geometrical parameters. Although this makes design more challenging, it also offers exciting opportunities to imbue robots with sensing, actuating, and interactive functionalities that are not accessible using conventional approaches.

### Beam-based structures

Slender beam elements are widely used for programming the deformation pattern of flexible metamaterials, because they can be easily manufactured via standard 3D printing approaches. By carefully arranging elastic beam elements, a variety of unusual mechanical behaviors can be achieved. For example, if beams are arranged in a re-entrant microstructure, then a negative Poisson's ratio emerges. This class of metamaterials, called auxetics, become counter-intuitively thicker when stretched and thinner when compressed. Combining auxetic and non-auxetic clutches simplified a soft robot's three-step inchworm locomotion using a single actuator rather than three independent ones (Fig. 1A) (3). Auxetic materials also allow surfaces that conform well to complex curved shapes (4). Elastic beams not only bend but also buckle under axial compression. This simple phenomenon may trigger homogeneous and reversible pattern transformations in metamaterials composed of regular arrays of elastic beams. Such reconfigurations can also be triggered by applying negative pressure and were exploited for soft machines that produced a range of useful motions even when actuated with a single pressure input (Fig. 1B) (5). Moreover, buckling of beams enabled large deformations at small forces suitable for artificial muscles (6). Last, elastic beams can snap between two stable configurations, which can be exploited to store and release elastic strain energy, as demonstrated by an untethered soft robot propelling itself in response to water temperature changes (Fig. 1C) (7). These examples highlight the potential of exploiting the nonlinear response of beams in soft robot design, but these systems will require materials that can withstand large deformations (~50% strain) without breaking and fatigue failure.

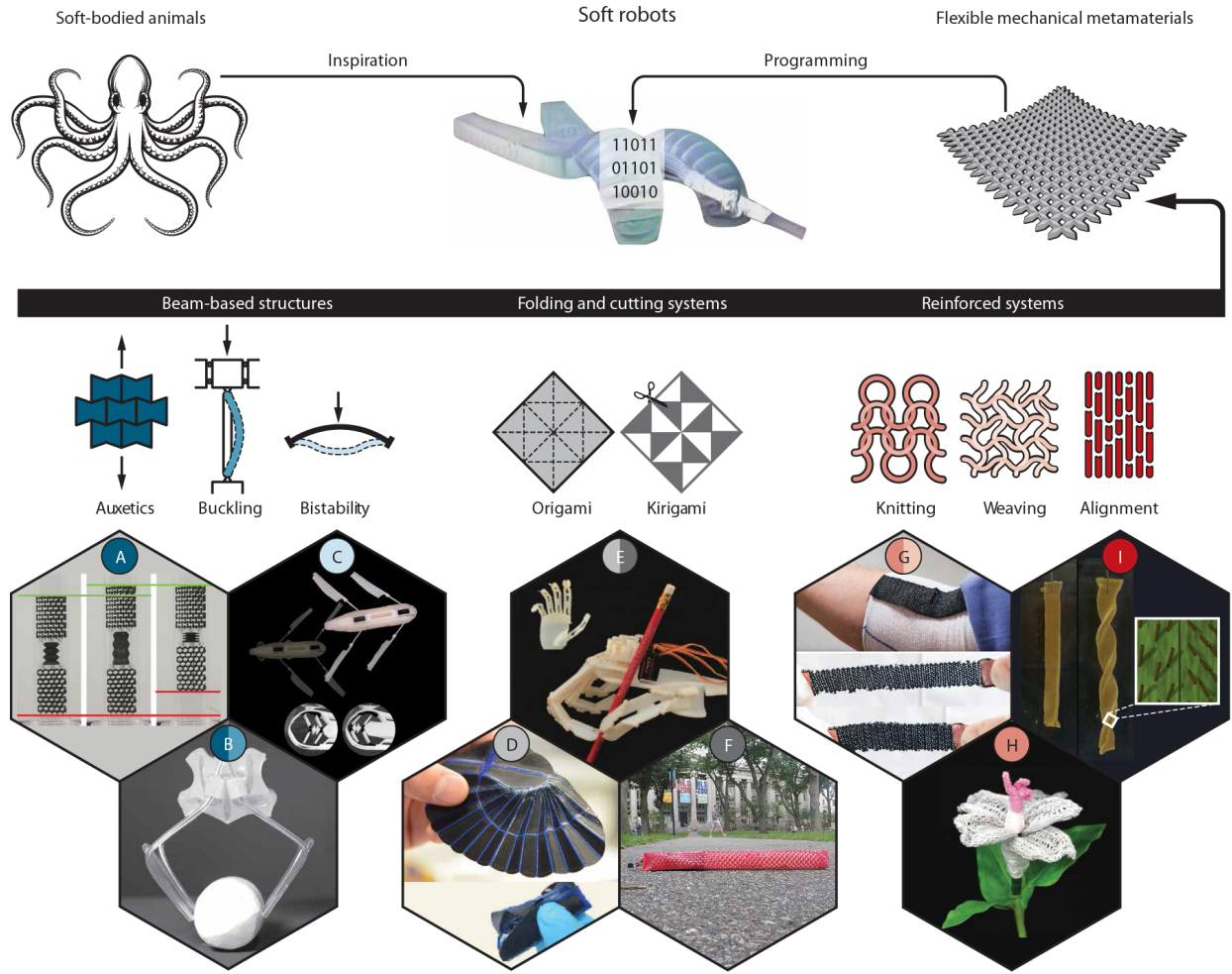

Fig. 1. **A metamaterial approach to soft robotics**. Robot responses (2) can be programmed into their squishy bodies by harnessing flexible metamaterials based on beams, origami and kirigami, and reinforced architectures. (A) Inchworm locomotion with a combination of auxetic and non-auxetic clutches (3). (B) Buckling-induced auxetic actuators (5). (C) Untethered propulsion with temperature-responsive bistable elements (7). (D) Bistable origami wing with programmed fast-morphing and load-bearing capabilities (9). (E) An origami-inspired foldable robotic hand (10). (F) Rectilinear locomotion enhanced with kirigami skin (11). (G) Knitted and woven artificial muscles (12). (Credit: T. Balkhed, Linköping University.) (H) A morphable knitted blooming flower triggered with SMA actuators (13). (I) Chiral composite actuators made of magnetically aligned particles in a swellable gelatin matrix (14).

### Origami and kirigami

The origami and kirigami arts of folding and cutting sheets to form 3D metamaterial objects have been explored for programmable morphing in robotic applications (8). Origami-inspired metamaterials, for example, can be created by folding thin sheets along predefined creases. However, conventional origami designs are not able to carry mechanical loads. Inspired by the programmable folding of the earwig wing, a bistable origami combined load-bearing capacity and fast-morphing capabilities through multimaterial 3D printing of rigid plates connected by soft stretchable joints (Fig. 1D) (9). In another study, the Miura-Ori folding template was used to create an origami-like robotic hand (Fig. 1E) (10). Kirigami structures were developed to enhance the crawling capability of soft actuators by mimicking the stretchability, friction anisotropy, and active anchorage of snake skins (Fig. 1F) (11). Although some structures can be easily fabricated by simple folding and laser-cutting approaches, more elaborate 3D metamaterial designs that can be integrated into other robotic parts will likely require multimaterial manufacturing technologies that offer shape complexity with a broader spectrum of material chemistries.

### Reinforced systems

Motion can also be programmed in reinforced systems through localized alignment of anisotropic building blocks along specific directions. Reinforced structures can be designed to undergo deformations from bending to twisting and coiling, and the stiffer nature of reinforcement can be beneficial in robotic parts subjected to higher mechanical loads. For these systems, actuation is often achieved by combining the anisotropic mechanical behavior of the reinforcement with a stimuli-responsive soft phase. Such materials can be realized using a variety of techniques, such as knitting, weaving, and alignment of particles in an external field. Knitted and woven fabrics have recently attracted metamaterial designers because of their potential for programming flexibility, strength, and shape change by tweaking the arrangements of yarn. Artificial muscles made of textile actuators or "textuators" were fabricated by knitting and weaving cellulose yarns dyed with an electro-responsive coating (Fig. 1G) (12), whereas morphing 3D objects were made by knitting hybrid fibers containing shape memory alloy (SMA) wires (Fig. 1H) (13). Inspired by local orientation of cellulose microfibrils in seed pods, swelling-induced bilayer composite actuators were fabricated by orienting magnetized alumina platelets within a gelatin matrix (Fig. 1I) (14). Although complexly shaped objects with deliberate reinforcement architectures have been demonstrated (15), the alignment, weaving, and knitting approaches for manufacture of reinforced metamaterials often require processing steps in addition to the usual casting and printing techniques for shaping.

## Perspectives

The performance of soft robots can be enhanced by exploring the ample design space offered by flexible metamaterials and devising manufacturing technologies that enable their physical realization. Building a massive library of designs from which functional modules can be identified and pieced together in flexible metamaterial architectures will enable soft robots that can not only morph and safely interact with their surroundings but also perform logical operations, sense, and adapt, even in unstructured environments. Bioinspiration and artificial intelligence (AI) algorithms are becoming powerful tools to explore the vast design space. In a top-down approach, bioinspiration offers abundant designs that have been perfected through natural evolution to perform targeted functions. The challenge is to break down these complex functions and reverse-engineer architectures to establish bioinspired designs that fulfill the desired tasks while respecting the boundary conditions imposed by the environment. From a bottom-up perspective, AI should guide our search for designs suitable for performing complex end functions. Once promising designs have been preselected, their practical implementation will require advanced technologies for manufacturing the idealized metamaterial architectures. Recent advances in 3D printing soft materials using multimaterial and directed-assembly approaches have shown potential to address the fabrication challenge. Through cross-interdisciplinary approaches involving materials science, mechanical engineering, AI, and robotics, flexible metamaterials will soon empower soft robots with currently unforeseen functionalities.

**Funding**: A.R. acknowledges support from the Swiss National Science Foundation under grant no. P3P3P2_174326. K.B. acknowledges support from the NSF under grant no. 1830896. A.R.S. thanks the financial support from ETH Zürich and from the Swiss National Science Foundation (Consolidator grant number BSCGI0_157696 and National Centre of Competence in Research Bio-Inspired Materials).

**Competing interests**: A.R.S. has cofounded the company Spectroplast, which is focused on 3D printing of silicones.

# References

[1] K. Bertoldi, V. Vitelli, J. Christensen, M. van Hecke, Flexible mechanical metamaterials, Nat. Rev. Mater. 2, 17066 (2015).

[2] R.F. Shepherd, F. Ilievski, W. Choi, S.A. Morin, A.A. Stokes, A.D. Mazzeo, X. Chen, M. Wang, G.M. Whitesides, Multigait soft robot, Proc. Natl. Acad. Sci. U.S.A 108, 20400-20403 (2011).

[3] A.G. Mark, S. Palagi, T. Qiu, P. Fischer, Auxetic metamaterial simplifies soft robot design, Proceedings of the 2016 IEEE International Conference on Robotics and Automation, 16 to 21 May 2016, Stockholm, Sweden, pp. 4951-4956.

[4] M. Konaković-Luković, J. Panetta, K. Crane, M. Pauly, Rapid deployment of curved surfaces via programmable auxetics, ACM Transactions on Graphics (TOG), 37 (4), No. 106, (2018).

[5] D. Yang, B. Mosadegh, A. Ainla, B. Lee, F. Khashai, Z. Suo and K. Bertoldi, G.M. Whitesides, Buckling of elastomeric beams enables actuation of soft machines, Adv. Mater. 27, 6323-6327 (2015).

[6] D. Yang, M.S. Verma, J-H. So, B. Mosadegh, C. Keplinger, B. Lee, F. Khashai, E. Lossner, Z. Suo, G. M. Whitesides, Buckling pneumatic linear actuators inspired by muscle, Adv. Mater. Technol. 1(3), 1600055 (2016).

[7] T. Chen, O.R. Bilal, K. Shea, C. Daraio, Harnessing bistability for directional propulsion of soft, untethered robots, Proc. Natl. Acad. Sci. U.S.A 115, 5698-5702 (2018).

[8] E. Hawkes, B. An, N.M. Benbernou, H. Tanaka, S. Kim, E.D. Demaine, D. Rus, R.J. Wood, Programmable matter by folding, Proc. Natl. Acad. Sci. U.S.A 107, 12441-12445 (2010).

[9] J. Faber, A.F. Arietta, A.R. Studart, Bioinspired spring origami, Science 359, 13861391 (2018).

[10] S. Kamrava, D. Mousanezhad, S.M. Felton, A. Vaziri, Programmable origami strings, Adv. Mater. Technol. 3, 1700276 (2018).

[11] A. Rafsanjani, Y. Zhang, B. Liu, S.M. Rubinstein, K. Bertoldi, Kirigami skins make a simple soft actuator crawl, Sci. Robotics 3, eaar7555 (2018).

[12] A. Maziz, A. Concas, A. Khaldi, J. Stålhand, N-K. Persson, E.W.H. Jager, Knitting and weaving artificial muscles, Sci. Adv. 3, e1600327 (2017).

[13] M-W. Han, S-H. Ahn, Blooming knit flowers: Loop-linked soft morphing structures for soft robotics, Adv. Mater. 29, 1606580 (2017).

[14] R.M. Erb, J.S. Sander, R. Grisch, A.R. Studart, Self-shaping composites with programmable bioinspired microstructures, Nat. Commun. 4, 1712 (2013).

[15] M. Schaffner, J.A. Faber, L. Pianegonda, P.A. Rühs, F. Coulter, A.R. Studart, 3D printing of robotic soft actuators with programmable bioinspired architectures, Nat. Commun. 9, 878 (2018).